# Colossal Magnetocapacitance Effect and Multiferroism of Polycrystalline La$_{0.2}$Pb$_{0.7}$Fe$_{12}$O$_{19}$

*Guo-Long Tan[*], Hao-Hao Sheng*
*State Key Laboratory of Advanced Technology for Materials Synthesis and Processing,*
*Wuhan University of Technology, Wuhan 430070, China*

## Abstract

The mutual control of the electric and magnetic properties of a multiferroic solid is of fundamental and great technological importance. We report here on the colossal magnetoelectric coupling effect of polycrystalline La$_{0.2}$Pb$_{0.7}$Fe$_{12}$O$_{19}$. A large classic ferroelectric hysteresis loop with full saturation and a strong magnetic hysteresis loop were observed simultaneously in polycrystalline La$_{0.2}$Pb$_{0.7}$Fe$_{12}$O$_{19}$ due to the coexistence of an off-centered FeO$_6$ octahedron in its unit cell and electron spins in the partially filled 3d orbits of the Fe$^{3+}$ ions. The coupling voltage and capacity demonstrate giant oscillations, along with magnetic field; the maximum magnetocapacitance ratio exceeds $1.90 \times 10^5$% at 80 Hz. The substitution of La ions with Pb ions progressively stabilized the conical spin structure, which gave rise to the spin-current induced capacity oscillations upon magnetic field.

## 1. Introduction

Multiferroics are materials which not only show ferroelectric and magnetic order simultaneously, but also display coupling between these orders *(1-3)*. These materials could produce new technologies in which the low power and high speed of field-effect electronics are combined with voltage-controlled ferromagnetism *(4, 5)*. Such ideal multiferroic materials still remain rare because of the mutual exclusivity of ferroelectricity and ferromagnetism under normal circumstances *(2,6-8)*. After long-term efforts, many single-phase multiferroic compounds have been discovered, such as BiFeO$_3$ *(9)* and RMnO$_3$ *(10)*, all of which exhibit simultaneous magnetic and electric order below the critical temperature *(11, 12)*. However, the properties of these few compounds are insignificant in comparison with those of useful ferroelectrics or ferromagnets, as their spontaneous polarizations or magnetizations are smaller by a factor of 1,000 or more *(4, 13)*. Recently numerous studies have focused particularly on the class of so-called magnetic ferroelectrics *(7, 14, 15)*, such as the perovskites $RMnO_3$ *(7)*, $MnWO_4$ *(16, 17)* and Y-type hexagonal ferrites *(18)*, in which ferroelectricity is induced by magnetic order. Although a dramatic variation of the electric polarization (P) with the magnetic field (B) in those compounds *(19)* might be useful for application, the phenomena occur mostly at low temperatures *(14, 15)*, and the magnetoelectric (ME) effect in such systems is often observed with field magnitudes of a few teslas and the related ME susceptibility is still too small *(20, 21)*. Meanwhile, electric polarization is two to three orders of magnitude smaller than that in typical ferroelectrics *(7, 8, 22)*. For applications, however, it will be necessary to generate and control the ME effects at room temperature and by low magnetic fields. Hence, it is a long standing challenge in the research of multiferroics to improve the operating temperature *(15, 22)* and the ME sensitivity *(23, 24)*.

---

[*] Corresponding author, Tel: 0086-27-87870271, Fax: 0086-27-87879468, email: gltan@whut.edu.cn

M-type lead hexaferrite ($PbFe_{12}O_{19}$), a traditional ferrite oxide with high saturation magnetization and large magnetic coercivity, has been reported to present large spontaneous electric polarization at room temperature *(25,26)*. Its intrinsic ferroelectricity has been verified by several electric evidences in the literature *(26)*. Because the room-temperature multiferroism is essential to the realization of multiferroic devices that exploit the coupling between ferroelectric and ferromagnetic orders at ambient conditions, the coexistence of ferroelectricity and ferromagnetism in pure $PbFe_{12}O_{19}$ is not enough *(26)*, and the realization of the ME coupling effect is more important for practical application in novel electric devices. However, the ME effect has not yet been achieved in pure $PbFe_{12}O_{19}$. We hope to obtain such an ME effect by replacing 3 Pb atoms with 2 La atoms while still keeping the balance of its valence charge. As such, $La_{0.2}Pb_{0.7}Fe_{12}O_{19}$ was conceived. We intend to obtain not only the multiferroic characteristics but also sufficient ME coupling effect in this new compound at room temperature. We expected that the new M-type hexaferrite $La_{0.2}Pb_{0.7}Fe_{12}O_{19}$ would demonstrate a resultant room-temperature ME effect. We will present the preparation process of polycrystalline $La_{0.2}Pb_{0.7}Fe_{12}O_{19}$, the simultaneous occurrence of large ferroelectricity and strong ferromagnetism, and its colossal magnetocapacitance effect at room temperature.

## 2. Results and Discussion

### 2.1 Structure and Electric Properties of $La_{0.2}Pb_{0.7}Fe_{12}O_{19}$ compound

The single-phase $La_{0.2}Pb_{0.7}Fe_{12}O_{19}$ powders were prepared by a polymer precursor method, in which a certain amount of powder was pressed into a pellet that was sintered into ceramic. After sintering, the ceramic was subsequently heat-treated in $O_2$ three times to remove the oxygen vacancies and transform $Fe^{2+}$ into $Fe^{3+}$ in polycrystalline $La_{0.2}Pb_{0.7}Fe_{12}O_{19}$. The technical details are provided in the Supplementary Materials.

*Figure 1*a shows the X-Ray diffraction (XRD) pattern of our $La_{0.2}Pb_{0.7}Fe_{12}O_{19}$ compound; the underneath diffraction lines colored red are the standard diffraction spectrum of $PbFe_{12}O_{19}$. It may be seen that all the diffraction peaks of $La_{0.2}Pb_{0.7}Fe_{12}O_{19}$ match well with the standard red lines of pure $PbFe_{12}O_{19}$, and there is no second ferrite phase or any other impurities. The charge balance has already been conserved. The lattice parameters of $La_{0.2}Pb_{0.7}Fe_{12}O_{19}$ are calculated to be a = 5.8764 Å × c = 23.0612 Å in the hexagonal structure with space group *P6₃/mmc*. The unit cell volume of $La_{0.2}Pb_{0.7}Fe_{12}O_{19}$, however, has been contracted by 0.77% after the replacement of 3 Pb with 2 La atoms since the radius of the La ion is smaller than that of the Pb ion. The overlap of the two diffraction patterns reveals that $La_{0.2}Pb_{0.7}Fe_{12}O_{19}$ shares the same crystal structure with $PbFe_{12}O_{19}$. Therefore, we may conclude that La atoms have successfully been incorporated into the crystal lattice of $PbFe_{12}O_{19}$ by replacing 20% of the Pb lattice sites and leaving another 10% of sites vacant.

*Figure 1*b exhibits the field-dependent susceptibility of $La_{0.2}Pb_{0.7}Fe_{12}O_{19}$. Its magnetic structure is quite different from that of other Y-type hexaferrites *( 24, 27)* or helimagnets *(20)*, which exhibit three plateaus and a short steep ascent along with B. $La_{0.2}Pb_{0.7}Fe_{12}O_{19}$ first displays a long steep ascent, then a narrow plateau, a steep descent, and finally, a slow descent. The first steep ascent could be assigned to the modified helix phase within 0 mT to 100 mT, the narrow plateau region between 150 mT to 250 mT to the intermediate-I magnetic phase, then it exhibits successive intermediate phases (II, III) and finally reaches a collinear ferrimagnetic phase around 1.4 T, which are similar to those reported in $Ba_2Mg_2Fe_{12}O_{22}$ *(24)*.

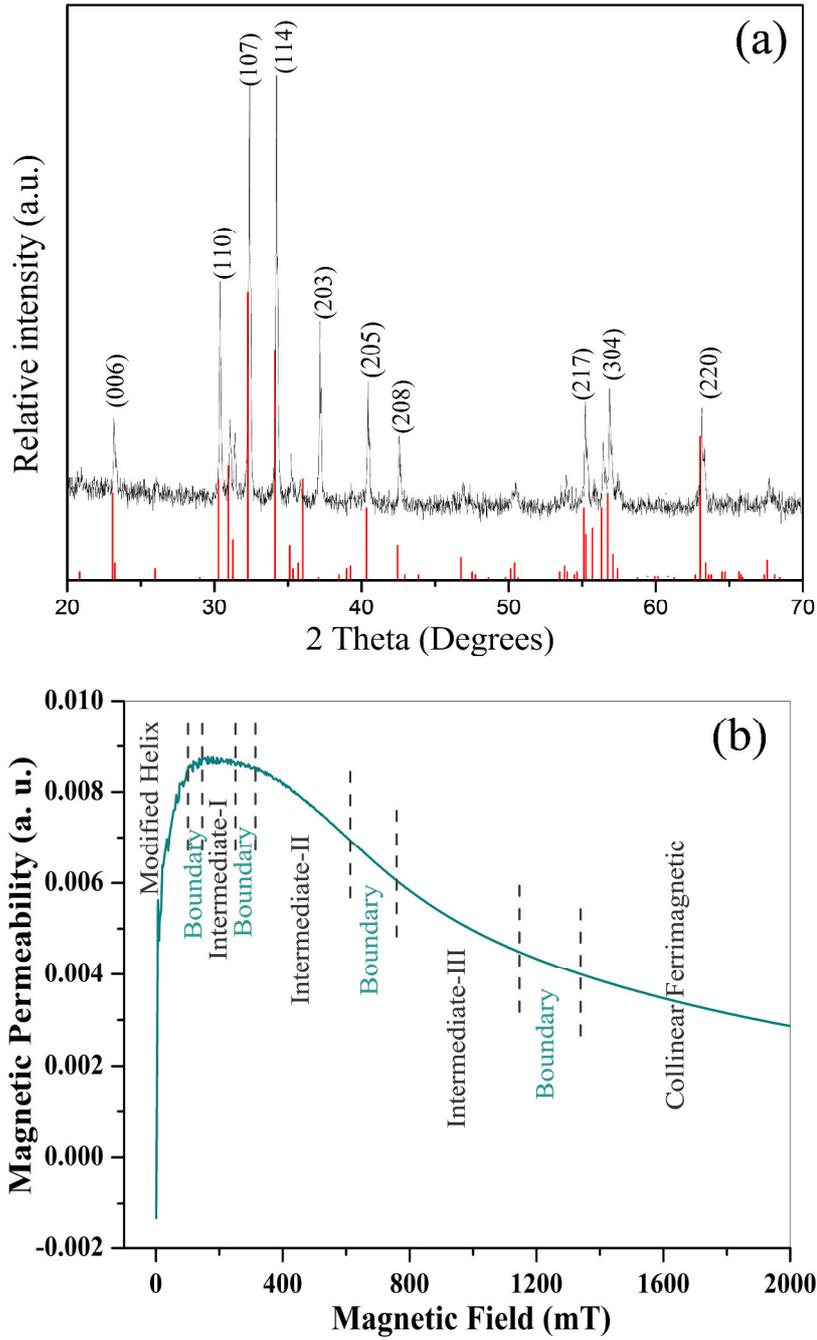

*Figure 1: (a) XRD pattern of the polycrystalline $La_{0.2}Pb_{0.7}Fe_{12}O_{19}$ with $O_2$ annealing process. The discrete red lines display the standard diffraction pattern of the $PbFe_{12}O_{19}$ compound (PDF#15-0623); (b) Field-dependent magnetic permeability ($\mu(B)$) of $La_{0.2}Pb_{0.7}Fe_{12}O_{19}$. The dashed lines represent the phase boundaries estimated for the down-sweep following the oscillation region of capacity as well as criteria in Ref. 27.*

## 2.2 Ferroelectric Properties of $La_{0.2}Pb_{0.7}Fe_{12}O_{19}$ Ceramics

We measured the complex impedance spectrum of polycrystalline $La_{0.2}Pb_{0.7}Fe_{12}O_{19}$ with and without annealing in $O_2$ (Figure S1 of Supplementary Materials). The resistivity estimated from the modules of the complex impedance is 1.8 M$\Omega$ and 5.4 k$\Omega$ for the samples with and without annealing in oxygen, respectively. This result indicates that the subsequent annealing in an $O_2$ atmosphere of the samples significantly affects the resistivity of $La_{0.2}Pb_{0.7}Fe_{12}O_{19}$. Hereinafter, we present only the results of electric and magnetic measurements as well as ME coupling data for the samples with subsequent annealing in oxygen.

For ferroelectric hysteresis loop (P-E) and dielectric measurement, both sides of the sample were coated with silver paste as electrodes. The temperature-dependent dielectric constant was also measured upon the poled $La_{0.2}Pb_{0.7}Fe_{12}O_{19}$ sample (Figure 2S). The result exhibits two dielectric anomalies located at 240 °C and 652 °C, respectively. The two dielectric anomalous peaks are similar to those of pure $PbFe_{12}O_{19}$ *(26)*; the first small peak is assigned to the transition from the ferroelectric to antiferroelectric phase, while the second tall peak is assigned to the transition from the antiferroelectric to paraelectric phase.

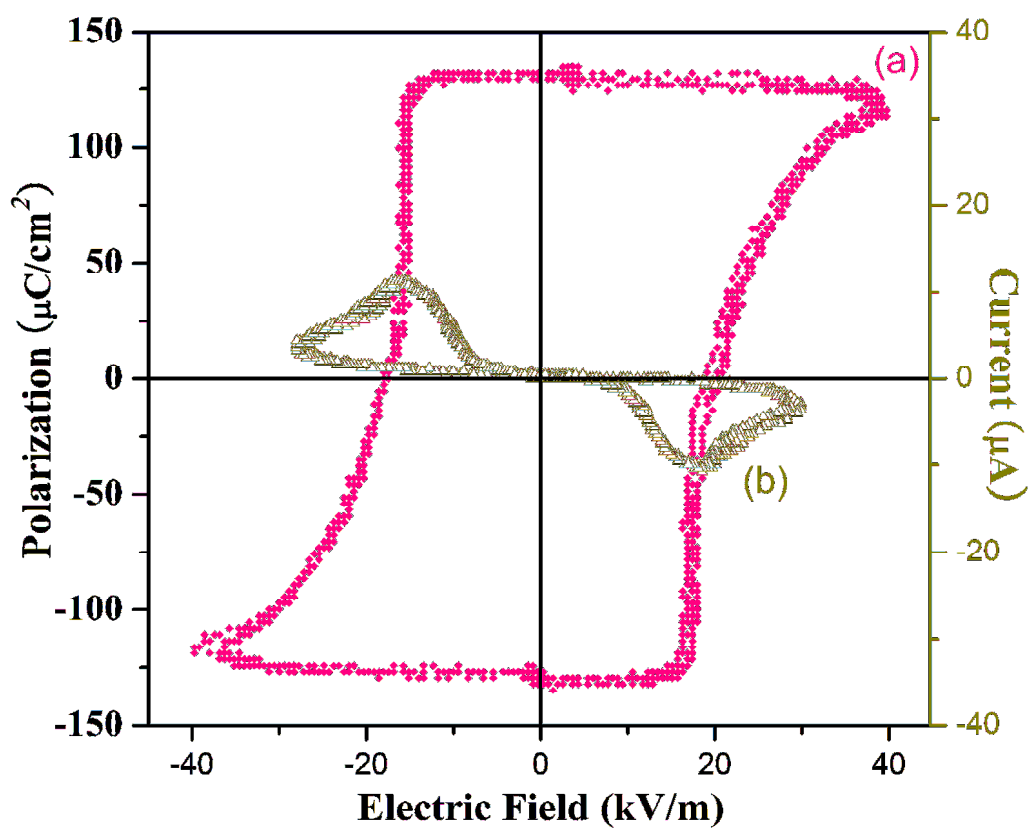

*Figure 2: (a) The ferroelectric hysteresis loop of polycrystalline $La_{0.2}Pb_{0.7}Fe_{12}O_{19}$ ceramic being measured at a frequency of 33 Hz and at room temperature; (b) the nonlinear I–V curve of $La_{0.2}Pb_{0.7}Fe_{12}O_{19}$ ceramic. The ceramic was sintered at 1000 ℃ for 1 h and subsequently heat-treated in an $O_2$ atmosphere for 9 h.*

The electric field-induced polarization performance was examined using a home-made ferroelectric measurement system termed ZT-IA at room temperature. The specimen was parallel-connected with a capacitor of 1 µF for compensation. The external electric field was input onto the sample by a triangular wave signal at the ultralow frequency of 33 Hz. *Figure 2*a shows a classic ferroelectric hysteresis loop of polycrystalline $La_{0.2}Pb_{0.7}Fe_{12}O_{19}$ with full saturation. In the hysteresis loop, a giant variation occurs when the external electric field is in the vicinity of the sample's coercive field. At the high electric field region, the polarization goes to saturation due to the even alignment of the electric orders along the direction of the electric field. The remnant polarization of the hysteresis loop is about 132.3 µC/cm$^2$. This value is higher than that of pure $PbFe_{12}O_{19}$ ceramic (104 µC/cm$^2$) *(26)* obtained by the same process. The coercive field of the polycrystalline $La_{0.2}Pb_{0.7}Fe_{12}O_{19}$ is 19.8 kV/m only.

*Figure 2*b demonstrates a nonlinear voltage–current (I–V) curve, which is similar to that of typical lead zirconate titanate (PZT) ceramics and provides us with additional evidence for the ferroelectricity of polycrystalline $La_{0.2}Pb_{0.7}Fe_{12}O_{19}$. When the ferroelectric polarization switches, the screening surface charges flow from one side of the electrode to the other, and an additional current is created momentarily, which creates two peaks at its coercive voltages. This result convinces us that the hysteresis loop indeed originates from polarization instead of linear current leakage. The classic ferroelectric hysteresis loop with full saturation, two nonlinear I–V peaks, and colossal variation of the dielectric constant near the Curie temperature offer enough evidence for us to verify that $La_{0.2}Pb_{0.7}Fe_{12}O_{19}$ is indeed a ferroelectric compound. The origin of the ferroelectricity of M-type lead hexaferrite has been discussed in detail in the previous literatures *(25, 26)*.

## 2.3 Magnetic Properties of the $La_{0.2}Pb_{0.7}Fe_{12}O_{19}$ Compound

Ferromagnetic characterization of polycrystalline $La_{0.2}Pb_{0.7}Fe_{12}O_{19}$ has been measured using the Physical Property Measurement System (PPMS) at room temperature. *Figure 3* demonstrates the magnetization hysteresis loop of $La_{0.2}Pb_{0.7}Fe_{12}O_{19}$. When the sample is placed into a uniform external magnetic field, the dipoles tend to align themselves along the field direction. As the external field strength increases, some dipoles will rotate to align with the external field, and thus the magnetic moment (M) increases with the field strength. Ultimately, all of the domains will have parallel magnetic moments. At this point, the magnetization of the sample will be at a maximum and will reach saturation. Reversal of the applied field will cause the magnetic moments to align in the opposite direction. The B-H path will therefore not follow the original path, but will trace a new path that 'lags behind' the old one (*Figure 3*). Ultimately, domain alteration will cease when all the magnetic moments are now parallel to the reversed applied field and saturation is again reached. In this way, the magnetization hysteresis loop is created (*Figure 3*). The saturated magnetization of polycrystalline $La_{0.2}Pb_{0.7}Fe_{12}O_{19}$ is estimated to be about 50 emu/g at H = 10 kOe. Its remnant magnetic moment ($M_r$) is about 28.7 emu/g and the coercive field ($H_c$) is around 3439.2 Oe. The remnant moment of $La_{0.2}Pb_{0.7}Fe_{12}O_{19}$ has been reduced by a small amount while its coercive field has been enhanced by 1114 Oe in comparison with pure $PbFe_{12}O_{19}$ *(26)*, since the substituted La provides additional unpaired f electrons. The thick hysteresis loop reflects the strongly magnetic feature of $La_{0.2}Pb_{0.7}Fe_{12}O_{19}$.

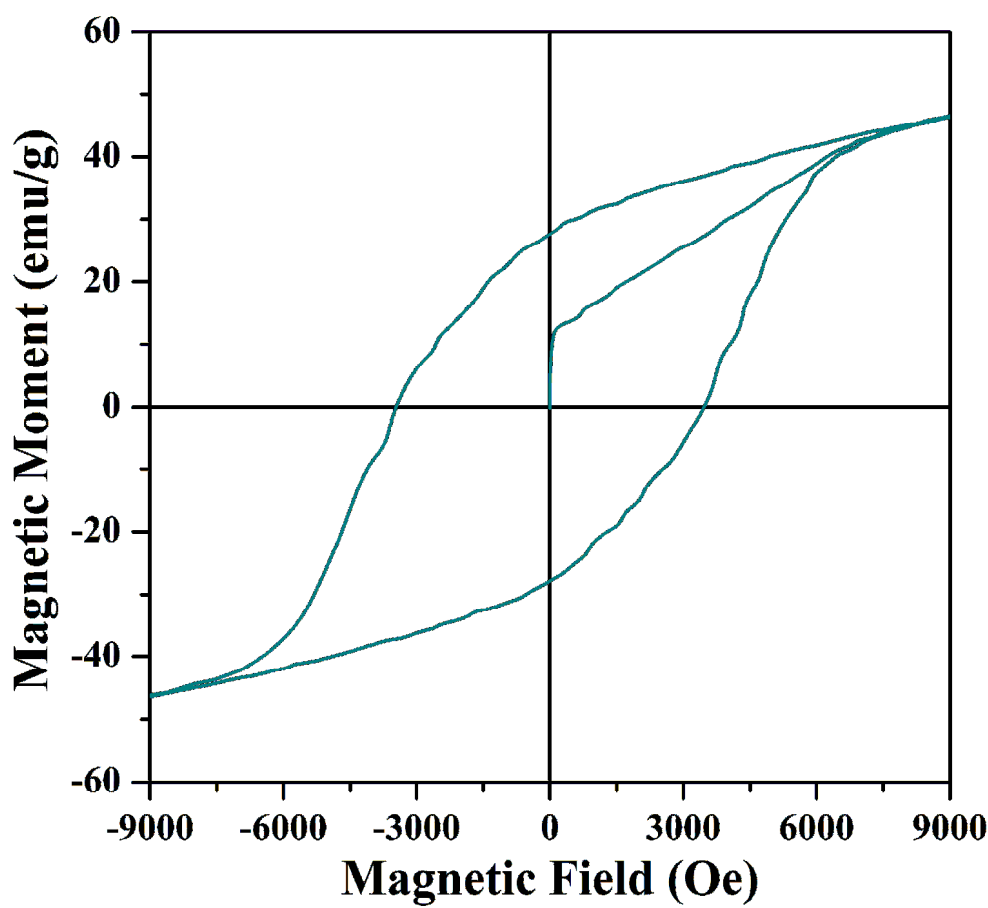

Figure 3: *Magnetic hysteresis loop of polycrystalline $La_{0.2}Pb_{0.7}Fe_{12}O_{19}$ being sintered at 1000 °C for 1 h and subsequently annealed in $O_2$ for 9 hs.*

## 2.4 Colossal Magnetocapacity of Polycrystalline $La_{0.2}Pb_{0.7}Fe_{12}O_{19}$

First of all, we need to determine if $La_{0.2}Pb_{0.7}Fe_{12}O_{19}$ could generate coupling charge upon the application of an external magnetic field. Under this consideration, we set up a simple system for the ME coupling measurement, which was performed by measuring the output coupling voltage (V) as a function of the magnetic field (B). The polycrystalline $La_{0.2}Pb_{0.7}Fe_{12}O_{19}$ ceramic was coated with silver electrodes on both sides and then placed in a space between two electromagnets. Upon the application of the magnetic field, the microvoltmeter (which was linked with the electrodes of the sample) would display the output coupling voltage. *Figure 5*a shows the variation of such ME coupling voltages as a function of B at room temperature. There is almost no coupling voltage output at zero B. By applying a low magnetic field within 8 mT to 100 mT, the coupling voltage appears and remains at the small constant value of 7.8 mV. With further increase of B from 100 mT to 200 mT, the coupling voltage shows a rapid enhancement from 7.8 mV to 47.6 mV, which then starts to rapidly decrease down to 22 mV when the magnetic field extends from 200 mT to 350 mT (*Figure 5*a). Afterwards, with further increase of the B up to 1000 mT, the coupling voltage remains almost unchanged. There is a kind of oscillation of the coupling voltage along with the magnetic field; the altitude of this oscillation is 510%, which reflects a strong ME coupling effect. The B-dependent output voltage originates from the spin current ($M_c$)-induced polarization.

Now we switch the electrode links of the sample from the microvoltmeter to a Wayne Kerr 6500B LCR Precision impedance analyzer to measure the variation of the capacitance with the applied magnetic field. *Figure 5*b displays a remarkably strong dependence of the capacity (or dielectric constant) on the applied magnetic field at low frequencies (f < 100 Hz). Its dielectric constant exhibits a remarkable change with the applied magnetic field and is very sensitive to the frequency. As B is below 100 mT, the dielectric constant shows almost no change. The incremental increase of B from 100 mT to 150 mT leads to a colossal enhancement of the dielectric constant from 23 up to 63813 at 80 Hz (*Figure 5*b). There are two huge oscillation peaks of the dielectric constant along with magnetic field centered at 200 mT and 350 mT, respectively. The amplitude of the oscillation is inversely proportional to the frequency; the lower is the frequency, the larger is the amplitude of oscillation. The maximum value of the magnetocapacitance ratio exceeds $1.1 \times 10^4$% at 50 Hz and $1.90 \times 10^5$% at 80 Hz, which are larger by more than three orders of magnitude than those previously reported (500% in $DyMnO_3$ *(10)*, 7% in $EuTiO_3$ *(28)* and 10% in $TbMnO_3$ *(7)*). It is noteworthy to mention that our colossal magnetocapcitance effect of $La_{0.2}Pb_{0.7}Fe_{12}O_{19}$ was realized at room temperature, while that of other reported compounds occurred at low temperature: DyMnO₃ at 18 K *(29)*, Ba₂Mg₂Fe₁₂O₂₂ at 5 K *(24)*, Ba₀.₅Sr₁.₅Zn₂(Fe₁₋ₓAlₓ)₁₂O₂₂ at 30 K *(10)*, and BaFe₁₂₋ₓScₓMg₆O₁₉ at 30 K *(30)*.

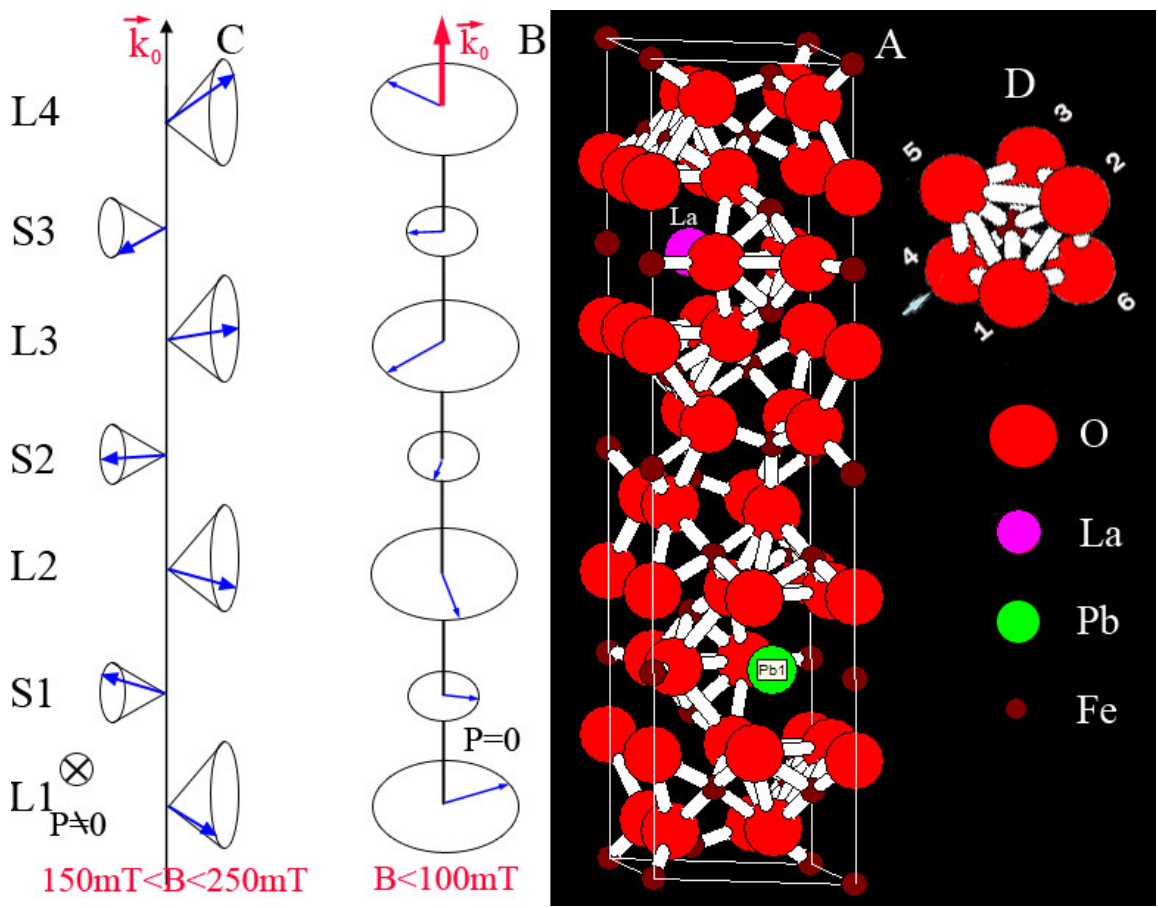

*Figure 4: Schematic crystal structure of La₀.₂Pb₀.₇Fe₁₂O₁₉ with space group P63/mmc. (A) The magnetic structure consists of alternative L and S blocks, corresponding to the octahedron and tetrahedron layers with large and small magnetic moments, respectively. The schematic magnetic structures display modified helix spins for the magnetic field region of (B) B < 100 mT and the cycloid conical spins for the region of (C) 150 mT < B < 250 mT. (D)A model of the off-centered FeO₆ octahedron is also illuminated.*

The colossal response of the dielectric constant to the low magnetic field in the vicinity of 150 mT at frequencies less than 100 Hz is likely associated with a magnetic phase transition. Such a phase transition is referred to the magnetic phase diagram of La₀.₂Pb₀.₇Fe₁₂O₁₉ in *Figure 1*b, where the helix spin structure exists in the first magnetic phase within the low B region of 0 mT to 100 mT, while the intermediate phase-I phase with cycloid heliconical spins resides within the B field region of 150 mT to 250 mT, similar to the magnetic phase structure of BaFe₁₂₋ₓScₓO₁₉ *(31, 32)*. The helix and conical spin structures are illustrated in *Figure 4*B & C, respectively, which are closely related to its layered crystal structure in *Figure 4*A. Within the low magnetic field region of 8 mT to 100 mT, $La_{0.2}Pb_{0.7}Fe_{12}O_{19}$ takes a modified helix state (*Figure 4*B), whose vector of $(\mathbf{S}_i \times \boldsymbol{S}_j)$ is parallel to $k_0$ ($e_{ij}$), and thus $P \propto \boldsymbol{e}_{ij} \times (\boldsymbol{S}_i \times \boldsymbol{S}_j) = 0$. In this case, no polarization could be induced and the coupling voltage inside this region is trivial, so the response of the dielectric constant to the magnetic field is dismissible (*Figure 5*a).

When B transverses consecutively from 100 mT to 150 mT, it comes across to the first magnetic phase boundary, where the helix spin and cycloid conical spin coexist (*Figure 1*b). In this region, the content of the cycloid conical spins starts to increase with B from 0% at 100 mT to 100% at 150 mT; the higher is the value of B, the greater is the concentration of the conical spins. The cycloid conical spins would align easily along B to form a transverse conical state (*Figure 4*B), which generates spin current ($M_c$) or coupling voltage (P) in the ab plane upon magnetic actuation. With gaining more and more cycloid conical spins in the first magnetic phase boundary, the spin-current induced coupling voltage (P) would increase rapidly with B (*Figure 5*a). The rapid growth of the ME coupling voltage will then induce a giant enhancement of the capacity or dielectric constant with B until 150 mT (*Figure 5*b). Further increasing B into the region of 150 mT to 250 mT, $La_{0.2}Pb_{0.7}Fe_{12}O_{19}$ will be in the intermediate-I phase which is composed of full cycloid conical spins (*Figure 1*b). Since the transverse conical spins have the highest coupling efficient and could induce the largest spin current or voltage in comparison with other magnetic structures, the capacity or dielectric constant remains at the maximum value with little variation inside this region. Afterwards, the dielectric constant starts to decreases rapidly when B goes into the second magnetic phase boundary between the intermediate-I phase and II phase (250 mT to 300 mT), where the content of cycloid conical spins decreases with the B field. The observed colossal magnetocapacitance effect induced at low B is thus proposed to be associated with a cycloid conical spin ground state. The La substitution progressively stabilizes such a cycloid conical state, as reported in $Ba_2Mg_2Fe_{12}O_{22}$ *(24)*.

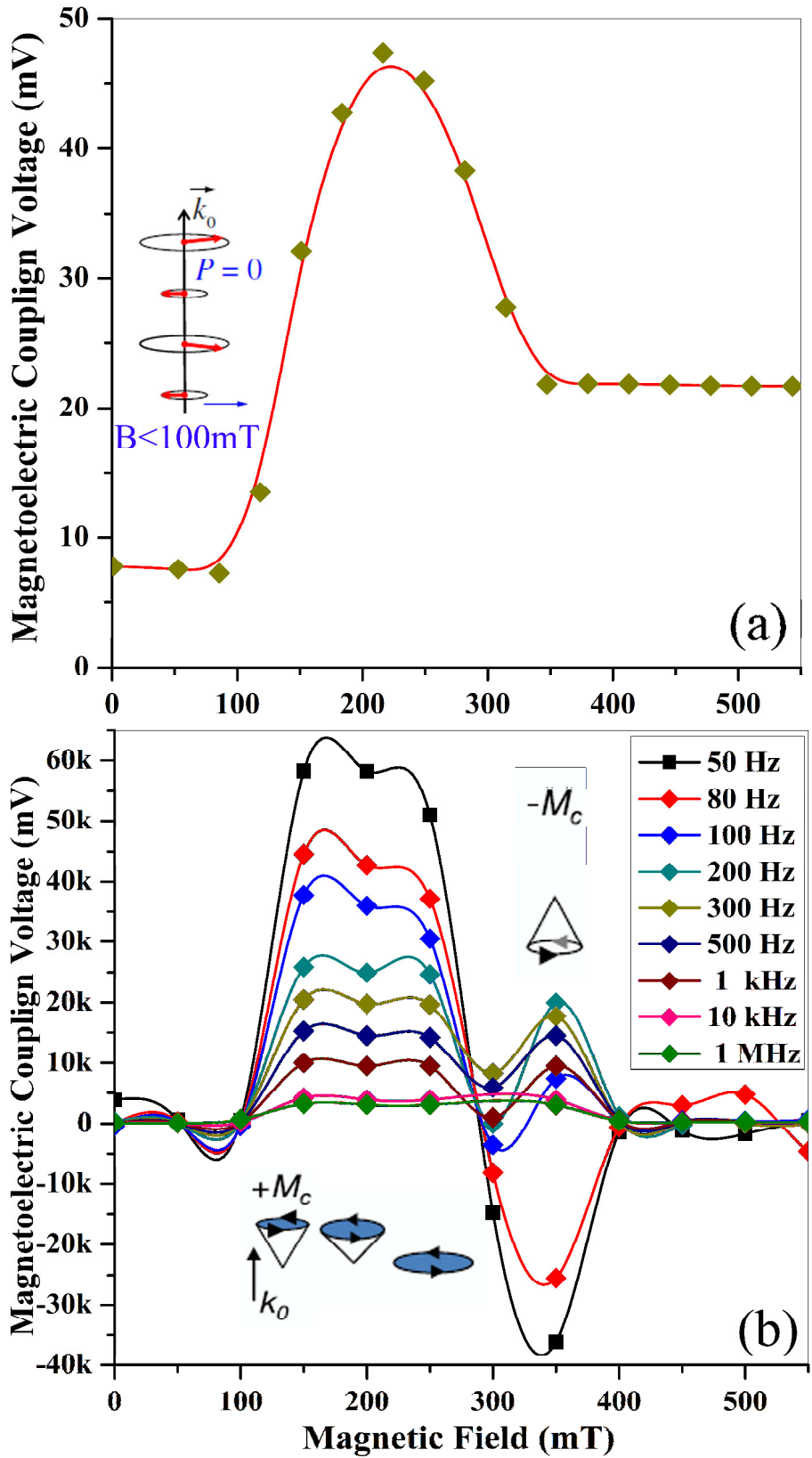

Figure 5: Magnetoelectric coupling effect of the polycrystalline $La_{0.2}Pb_{0.7}Fe_{12}O_{19}$; (a) magnetic-field dependent magnetoelectric coupling voltage. The inset shows the modified helix phase (20); (b) colossal magnetocapacitance effect: oscillation of relative dielectric constant (or capacity) with magnetic field for the polycrystalline $La_{0.2}Pb_{0.7}Fe_{12}O_{19}$ ceramics at room temperature. The insets display the cycloid conical spin structures (30) with positive $M_c$ (left) and negative $M_c$ (right).

The most interesting phenomenon is that the dielectric constant drops rapidly from positive to negative within the region of 300 mT to 350 mT, which is the stable region for the intermediate-II phase. We may attribute the sign change of the dielectric constant to the direction switch of the conical spin vector ($S_i \times S_j$) upon the tilting of B *(30)*, being similar to the mechanism for the negative sign of polarization (P) which appears in other M-type hexaferrites, $BaFe_{12-x}Sc_xMg_8O_{19}$ *(30)*. Supposing that P is initially positive, as B increases, the tilting angle of the cone axis monotonically decreases. As B traverses the field of 300 mT, the tilting angle of the cone becomes negative which causes the spin current ($M_c$) or voltage to reverse its direction and change the sign of P. The sign change of $(S_i \times S_j)_c$ switches the direction of P, which causes the dielectric constant to become negative upon the reversal of $M_c$ (right inset in *Figure 5*b).

The non-central symmetrical $FeO_6$ octahedron in the hexagonal unit cell was reported to be responsible for the electric polarization of $PbFe_{12}O_{19}$ *(20)*. According to the reported structural refinements in literatures, the $Fe^{3+}$ ions in the octahedra are indeed off-centered in M-type hexaferrites *(23, 27)*. $La_{0.2}Pb_{0.7}Fe_{12}O_{19}$ shares the same crystal structure with $PbFe_{12}O_{19}$, and thus should have the same genesis of ferroelectricity, which arises from its off-centered $FeO_6$ octahedron (*Figure 4*D), in which the $Fe^{3+}$ ion shifts away from the center of the octahedron while O shifts off its original corner position, resulting in large spontaneous polarization. In this way, large ferroelectricity and strong ferromagnetism are naturally merged together in $La_{0.2}Pb_{0.7}Fe_{12}O_{19}$, due to the coexistence of the off-centered $FeO_6$ octahedron in its unit cell and electron spins in partially filled 3d orbits of the Fe ions as well as the 3f orbits of lanthanum. Thus, the mutually exclusive electric and magnetic orders are syncretized in one single phase $La_{0.2}Pb_{0.7}Fe_{12}O_{19}$. Therefore, large ferroelectricity, strong ferromagnetism, and a giant ME coupling effect are all realized in one single phase $La_{0.2}Pb_{0.7}Fe_{12}O_{19}$ compound at room temperature.

## 3. Conclusion

In summary, our work directly presents the room-temperature multiferroic characterization and colossal magnetocapacitance effect of a modified M-type lead hexaferrite ($La_{0.2}Pb_{0.7}Fe_{12}O_{19}$) obtained from the method of atom substitution. The polycrystalline $La_{0.2}Pb_{0.7}Fe_{12}O_{19}$ sample demonstrates a classical polarization hysteresis loop with full saturation, two particular nonlinear I–V peaks, and dielectric anomalies near the Curie temperature, all of which verify its intrinsic ferroelectricity. The large magnetic hysteresis loop was also observed in polycrystalline $La_{0.2}Pb_{0.7}Fe_{12}O_{19}$, indicating its strong ferromagnetism. Furthermore, the ME coupling voltage and capacity (dielectric constant) exhibit giant oscillations along with magnetic field, and the maximum magnetocapacitance ratio exceeds $1.90 \times 10^5$%, which is larger by more than three orders of magnitude than those of other compounds previously reported. The cycloid conical spin is proposed to be responsible for the B induced spin current or voltage (P), which actuates a huge response of the capacity to B and thus gives rise to the colossal magnetocapacitance effect.

**Acknowledgements:**

The authors acknowledge the financial support from Hubei Natural Science Foundation, under the contract No. 2014CFB166; Open fund of State Key Laboratory of Advanced Technology for Materials Synthesis and Processing (Wuhan University of Technology) under the contract No. 2016-KF-15.

# Supply Materials

## ● Materials and Methods:

The single-phase $La_{0.2}Pb_{0.7}Fe_{12}O_{19}$ powders were prepared by a polymer precursor method, certain amount of power was pressed into a pellet which was sintered into ceramic. After sintering, the ceramic was subsequently experienced heat-treatment in $O_2$ three times to remove the oxygen vacancies and transform $Fe^{2+}$ into $Fe^{3+}$ in the $La_{0.2}Pb_{0.7}Fe_{12}O_{19}$ ceramic. Firstly, 0.4193g lead acetate trihydrate ($Pb(CH_3COO)_2 \cdot 3H_2O$) was dissolved in 15mL glycerin forming a clear solution, which was distilled in a rotary evaporator at 120℃ for 1h to remove the molecules of water. Meanwhile, 0.1367g Lanthanum nitrate ($La(NO_3)_3$) was dissolved in 3mL distilled water. The two solutions were stored in two small glass bottles respectively. Then we moved these glass bottles into a glove-box, where the following experimental process was performed to prevent ferric acetylacetonate from hydrolyzing. 5.2976g ferric acetylacetonate ($C_{15}H_{21}FeO_6$) was dissolved in a mixture solution of 100mL anhydrous ethanol and 50mL acetone in a 250mL three-neck flask. Afterwards the distilled lead acetate solution and the lanthanum nitrate solution were added into the ferric acetylacetonate solution, the obtained mixture solution was stirred at 70℃ for more than 8h. The obtained solution was assigned as precursor solution. Here, the molar ratio of atomic La+Pb to atomic Fe was set to 1:9.5 to balance the Pb loss during the subsequent heat treatment process. 1.0g polyethylene glycol was dissolved in a mixture solution of 5mL distilled water and 45mL ammonia, which was added into the above precursor solution. The final mixture solution was stirred at 70℃ for 24h to form a dispersion solution. The dispersion was then moved out of the glove-box and then centrifuged at a speed of 12000rpm for 15min to remove the water and organic molecules. The sediment was calcined at 450℃ for 1h, thus nanophase $La_{0.2}Pb_{0.7}Fe_{12}O_{19}$ powders were prepared. The powder was further calcined at 800℃ for another hour to fully remove the organic molecules. After grinding for 1 hour, 0.050g $La_{0.2}Pb_{0.7}Fe_{12}O_{19}$ powder was pressed into a pellet sample, which was sintered for 1h at 1000℃ in $O_2$ atmosphere into a polycrystalline ceramic specimen. The ceramic pellet was then annealed at 700℃ in $O_2$ atmosphere for 3h, then turned over for another 3hs heat treatment in oxygen atmosphere. Once again, the polycrystalline

$La_{0.2}Pb_{0.7}Fe_{12}O_{19}$ pellet was annealed at 600℃ for final 3h in $O_2$ atmosphere to remove the oxygen vacancies and transform $Fe^{2+}$ into $Fe^{3+}$ in polycrystalline $La_{0.2}Pb_{0.7}Fe_{12}O_{19}$ pellet to the utmost extent. Phase identification of the $La_{0.2}Pb_{0.7}Fe_{12}O_{19}$ power was performed by X-ray powder diffraction (XRD) using Cu-$K_\alpha$ radiation. As for P-E hysteresis loop measurement, both sides of the ceramic surfaces were coated with silver paste as electrode which was heat treated at 820℃ for 15 min. Then the ferroelectric hysteresis loop was measured upon the polycrystalline $La_{0.2}Pb_{0.7}Fe_{12}O_{19}$ pellet with electrode by using an instrument referred as ZT-IA ferroelectric measurement system. The temperature dependent dielectric property was measured upon the pellet using a LCR instrument (HP4284A). Magnetization measurement was carried out on a Quantum Design physical property measurement system (PPMS-9). Its complex impedance spectrum was measured upon a electrochemical station (Chenghua) within the frequency range of 0.1 Hz-1 MHz. The magnetocapacitance parameters of the polycrystalline $La_{0.2}Pb_{0.7}Fe_{12}O_{19}$ pellet were measured using a Wayne Kerr 6500B LCR station by applying a variable magnetic field.

We measured the complex impedance spectrum of polycrystalline $La_{0.2}Pb_{0.7}Fe_{12}O_{19}$ ceramics with and without annealing in $O_2$ (Figure S1 of supplementary Materials). The resistivity estimated from the modules of the complex impedance is 5.4k$\Omega$ and 1.8M$\Omega$ for the samples sintered with and without annealing in oxygen, respectively. This result indicates that the subsequent annealing in $O_2$ atmosphere of the samples significantly affects the resistivity of $La_{0.2}Pb_{0.7}Fe_{12}O_{19}$. The increase of the resistivity for M-type hexaferrites ($PbFe_{12}O_{19}$) annealing in oxygen has been reported previously *(1-3)*. The resistivity of polycrystalline $La_{0.2}Pb_{0.7}Fe_{12}O_{19}$ ceramics is around 2 orders lower than that of polycrystalline $PbFe_{12}O_{19}$ ceramics *(3)* due to the additional 10% lattice vacant sites after substitution of Pb with La. The subsequent heat-treatment in oxygen probably removes most of the oxygen vacancies and transforms the $Fe^{2+}$ into $Fe^{3+}$ ions and then reduces the current leakage and hopping of electrons between $Fe^{2+}$ and $Fe^{3+}$ ions 错误！未定义书签。. Therefore the resistivity of the samples has been greatly improved after annealing in $O_2$ atmosphere.

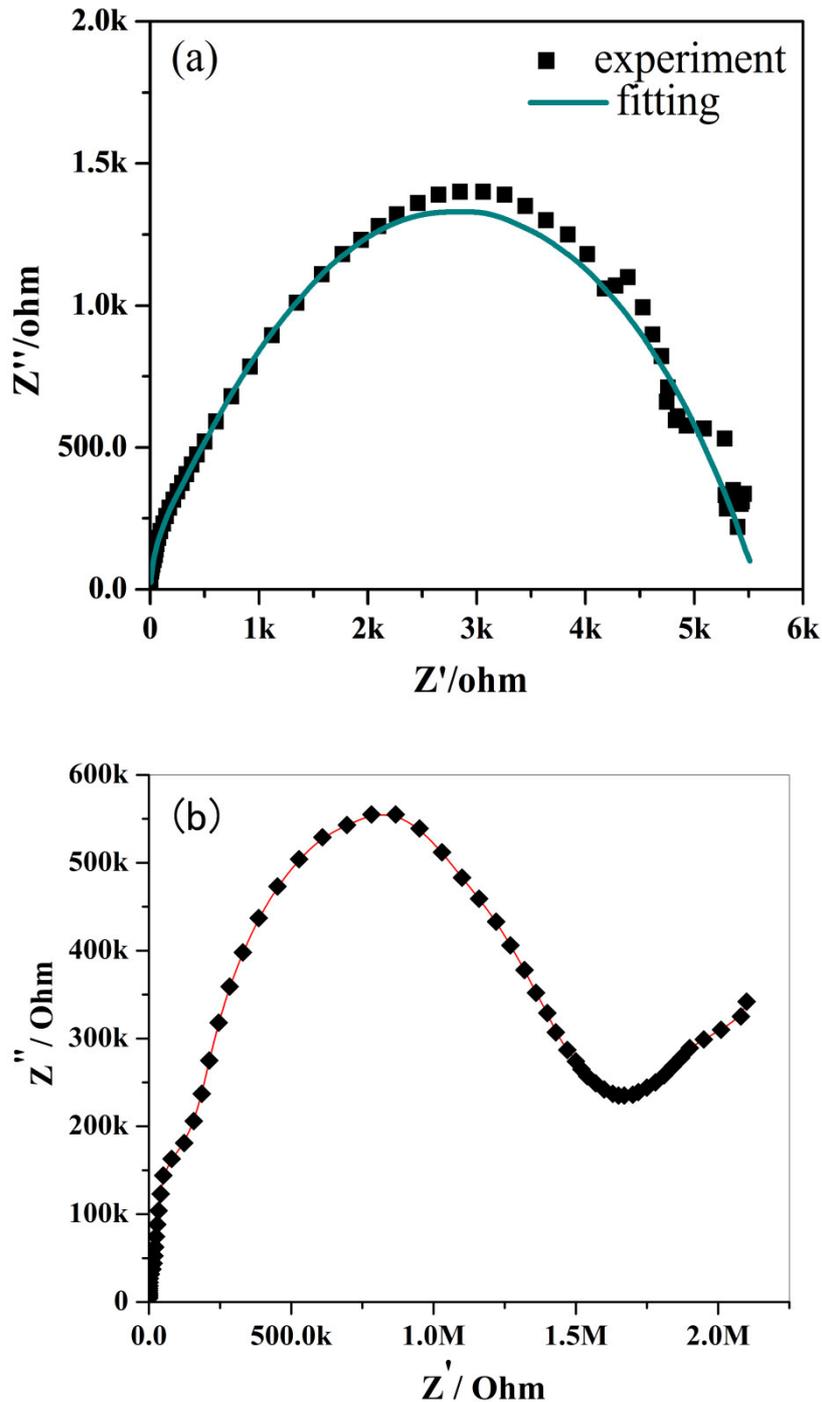

Figure 6S: *Complex impedance spectrum of La$_{0.2}$Pb$_{0.7}$Fe$_{12}$O$_{19}$ in the frequency range of 0.01 Hz to 1 MHz, (a) sintered at 1000℃ in air for 1 hour only; (b) sintered at 1000℃ in air and subsequently heat-treated in O$_2$ (both sides at 700℃ and one side at 600℃) for 9 hs total.*

      Figure 7S shows the temperature-dependent dielectric constant of poled La$_{0.2}$Pb$_{0.7}$Fe$_{12}$O$_{19}$ ceramic, the inset exhibits the variation of the dielectric constant as a function of temperature (<300℃) at different frequencies. At the measurement

frequency of 5 kHz, two dielectric anomalies are clearly seen, one is a small peak locating at 240°C, another one is a maximum dielectric anomaly peak at 652°C. The two dielectric anomaly peaks are similar to that of pure $PbFe_{12}O_{19}$ *(3)*, therefore the first small peak could be assigned to the transition from ferroelectric to antiferroelectric phase, while the second tall one to the transition from antiferroelectric to paraelectric phase. The temperature of the antiferro- to para-electric transition has been improved by 134°C in comparison with pure $PbFe_{12}O_{19}$ (518°C) *(3)* due to the substitution of La with Pb atoms. The first ferro- to antiferro- transition peaks are insensitive to frequency (inset of Figure 7S), indicating that it's not a diffuse phase transition, which is different from the relaxor characterization of pure $PbFe_{12}O_{19}$ *(3)*

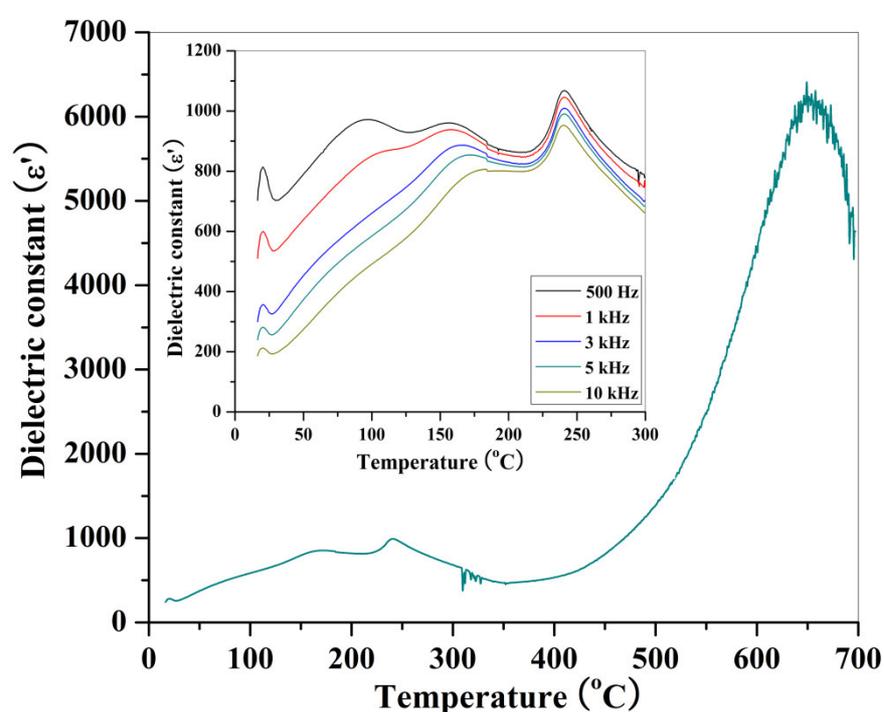

Figure 7S: *The variation of dielectric constant as a function of temperature for polycrystalline $La_{0.2}Pb_{0.7}Fe_{12}O_{19}$, the inset is the temperature dependence of the dielectric constant at different frequencies.*

Figure 8S shows the magnetic susceptibility (μ) of polycrystalline $La_{0.2}Pb_{0.7}Fe_{12}O_{19}$ measured at 250 mT as a function of temperature. The magnetic permeability of polycrystalline $La_{0.2}Pb_{0.7}Fe_{12}O_{19}$ at room temperature is 0.0114. The temperature dependence of permeability exhibits two clear sudden drops at 740K ($T_S$) and 861k ($T_C$) respectively. The two magnetization anomalies at 740K and 861K are likely associated with a reorientation transition from the phase of cycloid spiral spin to a collinear ferrimagnetic state and another one from ferrimagnetic state to paramagnetic phase, respectively, which is similar to that of Y-type hexaferrite: $Sr_3Co_2Fe_{24}O_{41}$ *(4)*. Therefore, the first peak ($T_s$) is corresponding to the magnetic phase transition, while the second one ($T_c$) is for the Curie temperature, at which point the magnet is changed into paramagnet. The Curie temperature of $La_{0.2}Pb_{0.7}Fe_{12}O_{19}$ has been improved by 135 K in comparison with that of pure $PbFe_{12}O_{19}$, which would be ascribed to the substitution of Pb with La atoms.

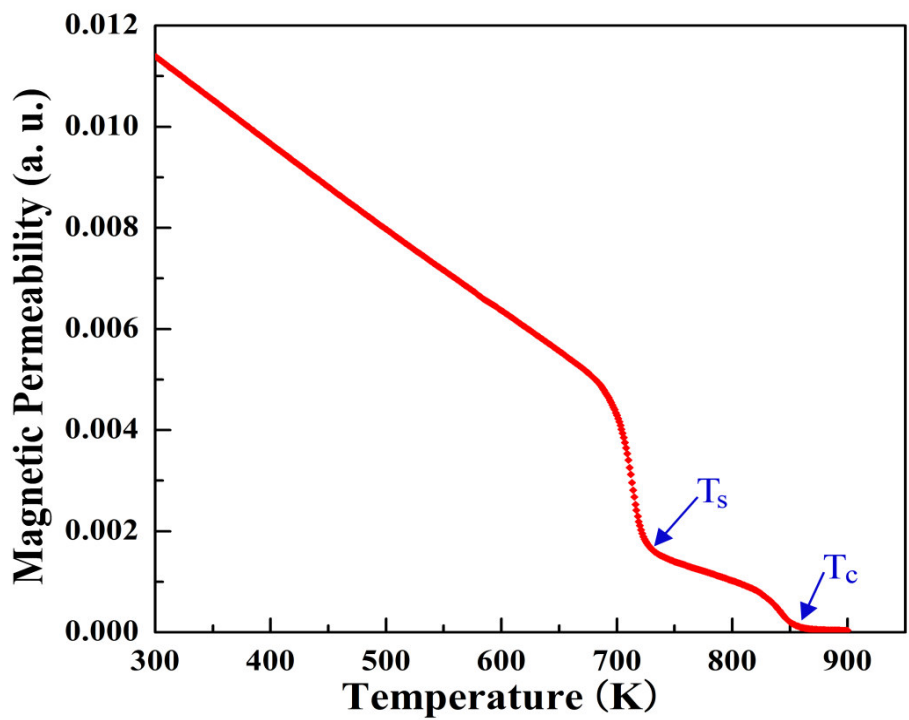

Figure 8S: *Temperature-dependent magnetic susceptibility (μ(T)) measured at B=250 mT,*

The M-type hexaferrite crystallizes in a hexagonal structure with 64 ions per unit cell on 11 different symmetry sites (*P*63/*mmc* space group), the diagram of its crystal structure is shown in Figure 9S. The 24 $Fe^{3+}$ atoms are distributed over five distinct sites: three octahedral sites (12*k*, 2*a* and 4*f2*), one tetrahedral (4*f1*) site and one hexahedral (trigonal bipyramidal) site (2*b*). $Fe^{3+}$ ions in 12*k*, 2*a*, and 2*b* sites (16 total per unit cell) have their spins up, while the $Fe^{3+}$ ions in 4*f1* and 4*f2* sites (8 total per unit cell) have their spins down (Figure S2), which results in a net total of 8 spins up, and therefore, a total moment of (8 x 5)μB = 40μB per unit cell. This explains why $La_{0.2}Pb_{0.7}Fe_{12}O_{19}$ compound demonstrates good magnetic properties with large $M_r$ (remnance) and $H_c$ (coercive field) value. The origin of the magnetic coupling in $La_{0.2}Pb_{0.7}Fe_{12}O_{19}$ is resulting from the super-exchange interactions between the three parallel (2*a*, 12*k* and 2*b*) and two antiparallel (4*f1* and 4*f2*) sublattices through the $O^{2-}$ ions.

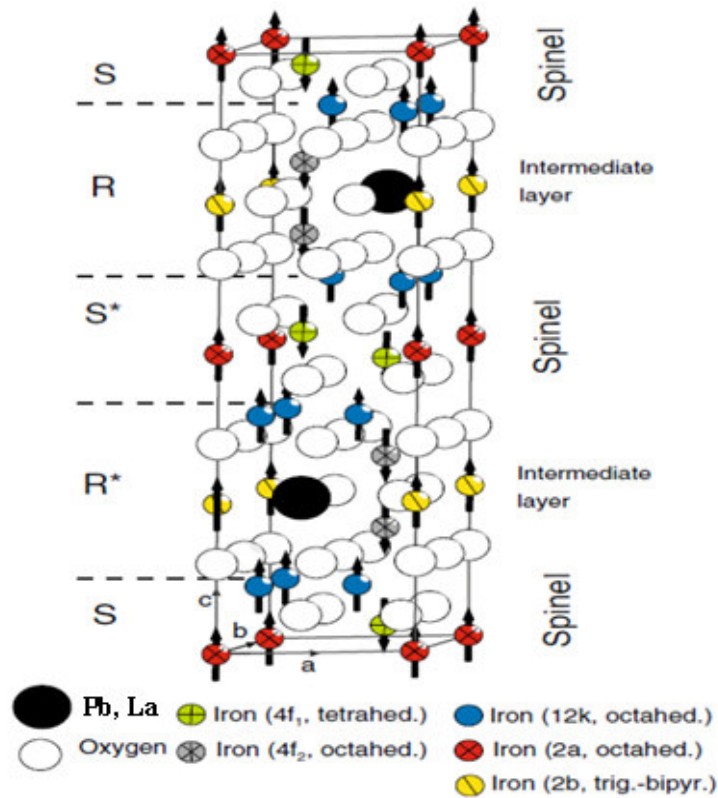

Figure 9S: *Crystal Structure Of M-type Hexaferrite showing a unit cell and position of ionic sites in the four spinel blocks (SRS\*R\*) and relative orientation of magnetic moments of $Fe^{3+}$ ions (5).*